\documentclass{ecctd99}
\usepackage[utf8]{inputenc}
\usepackage{graphicx}
\usepackage[english]{babel}
\usepackage{amsmath}
\usepackage{mathptmx}
\usepackage{etoolbox}
\usepackage{cite}
\usepackage{siunitx}
\usepackage[svgnames]{xcolor}
\usepackage[pdftex, unicode, bookmarks=false, bookmarksnumbered=true,
  pdfpagemode={UseOutlines}, plainpages=false, pdfpagelabels=true,
  colorlinks=true, linkcolor={DarkBlue}, citecolor={DarkBlue}, 
  urlcolor={DarkBlue}]{hyperref}
\usepackage[final]{microtype}
\usepackage[LGR,T1]{autofe}

\newcommand{\mailto}[1]{\href{mailto:#1}{\nolinkurl{#1}}}
\newcommand{\doi}[1]{\href{http://dx.doi.org/#1}{\nolinkurl{#1}}}

\renewcommand\normalsize{\fontsize{9.3}{11.3}\selectfont}

\makeatletter
\patchcmd{\maketitle}{%
  \twocolumn[\@maketitle]%
}{%
  \def\@makefntext##1{\parindent 1em\noindent ##1}%
  \let\@makefnmark\relax
  \twocolumn[\@maketitle]%
}{}{}
\makeatother

\newcommand{\SSI}{S\textsuperscript{2}I}
\newcommand{\Veff}[1][{}]{V_{\text{eff}_{#1}}}
\newenvironment{compactenumerate}{%
  \begin{list}{%
      \textit{\roman{enumi}.\quad}}{%
      \setlength{\topsep}{1ex}%
      \setlength{\leftmargin}{0pt}%
      \setlength{\parsep}{0pt}%
      \setlength{\itemsep}{0pt}%
      \setlength{\labelwidth}{0pt}%
      \setlength{\labelsep}{0pt}%
      \setlength{\itemindent}{0pt}      
      \usecounter{enumi}}}{%
  \end{list}}

\date{}

\title{%
  A Tailed Tent Map Chaotic Circuit Exploiting\\
  \SSI\ Memory Elements}

\author{S. Callegari$^*$, R. Rovatti$^*$, G. Setti$^{**}$\\
  {\normalsize $^*$DEIS - Università di Bologna, viale
    Risorgimento 2, 40136 Bologna - ITALY}\\
  {\normalsize e-mail: \mailto{scallegari@deis.unibo.it} |
    \mailto{rrovatti@deis.unibo.it}}
  \\
  {\normalsize $^{**}$DI - Università di
    Ferrara, via Saragat 1, 44100 Ferrara - ITALY} \\
  {\normalsize e-mail: \mailto{gsetti@ing.unife.it}}%
  \thanks{This is a pre-print version of a paper presented at the 1999
    European Conference on Circuit Theory and Design (ECCTD '1999).
    Published paper available in ``ECCTD '99: Proceedings of the 1999
    European Conference on Circuit Theory and Design'', European Circuit
    Society (ECS) 1999. Cite as:\protect\\[1ex]
    S.~Callegari, R.~Rovatti, G.~Setti, ``A tailed tent map chaotic
    circuit exploiting S\textsuperscript{2}I memory elements'', in
    Proc. of ECCTD’99, vol. 1, Stresa, IT, Sep. 1999,
    pp. 193–196.}%
}

\hypersetup{%
  pdftitle={%
     A Tailed Tent Map Chaotic Circuit Exploiting
  S²I Memory Elements}, 
  pdfauthor={%
    Sergio Callegari,
    Riccardo Rovatti,
    Gianluca Setti}}

\begin{document}
\maketitle
\thispagestyle{empty}

\begin{abstract}
  In the implementation of discrete time chaotic systems, designers
  have mostly focused on the choice and synthesis of suitable
  nonlinear blocks, while correct and fast operation cannot neglect
  analog memory elements.  Herein, a realization exploiting the Tailed
  Tent Map coupled with \SSI\ sample and hold stages is proposed.
  Interfacing problems between the memory and the processing
  sub-circuits are tackled and the possibility to obtain a fivefold
  speed improvement with respect to previously reported results is
  demonstrated by means of SPICE and other computer aided simulations.
\end{abstract}

\section{Introduction}

\begin{figure}[ht]
  \centering
  \resizebox{0.7\linewidth}{!}{\includegraphics{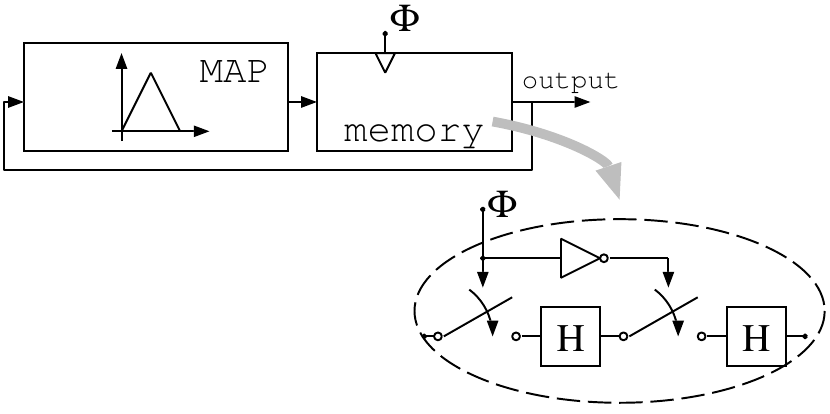}}
  \caption{%
    \label{fig:chaosloop}
    Basic scheme for a one-dimensional chaotic system. A memory
    element (cascade of two sample-and-hold stages) maintains the input of the
    map circuit stable during signal processing.}
\end{figure}

Several applications of chaotic
hardware (like in secure communication schemes \cite{Hasler:ISCAS-1994}, noise
generation, stochastic neural models \cite{Clarkson:WNNW-1993-18}, EMI
reduction, etc.\@) can exploit even the simplest kind of chaotic
systems, namely a one-dimensional map
(Figure~\ref{fig:chaosloop}).%
Although for obtaining chaotic behavior this arrangement requires a
fully \emph{analog} signal path, this does not prevent practical and
compact CMOS realizations
\cite{Delgado:EL-1993-25-2190,Langlois:NDES-1995,Callegari:ISCAS-1997}.
Indeed, interest in these systems has recently grown, mainly
focusing on how to obtain useful statistical properties, on the design
strategies, and on robustness issues \cite{Callegari:ISCAS-1997}. In
other words, the attention has been mainly devoted to the choice of
suitable nonlinear maps and on their synthesis.

However, for truly effective operation, \emph{all} the elements in the
signal path (thus including the analog memory in the feedback loop)
must be optimized having the requirements of a chaotic system in mind.

Recently, investigation in this direction has been proposed for a
representative set of systems defined by piecewise-affine (PWA) maps
and characterized by a uniform invariant probability density function
(PDF) of the generated samples~\cite{Callegari:NOLTA-1998}.  Two
results of such investigation are that even maps which are
traditionally considered robust with respect to implementation
inaccuracies suffer from S/H errors and that, among all these errors,
signal dependent ones seem to be the most critical.

\begin{figure}
  \resizebox{0.32\linewidth}{!}{\includegraphics{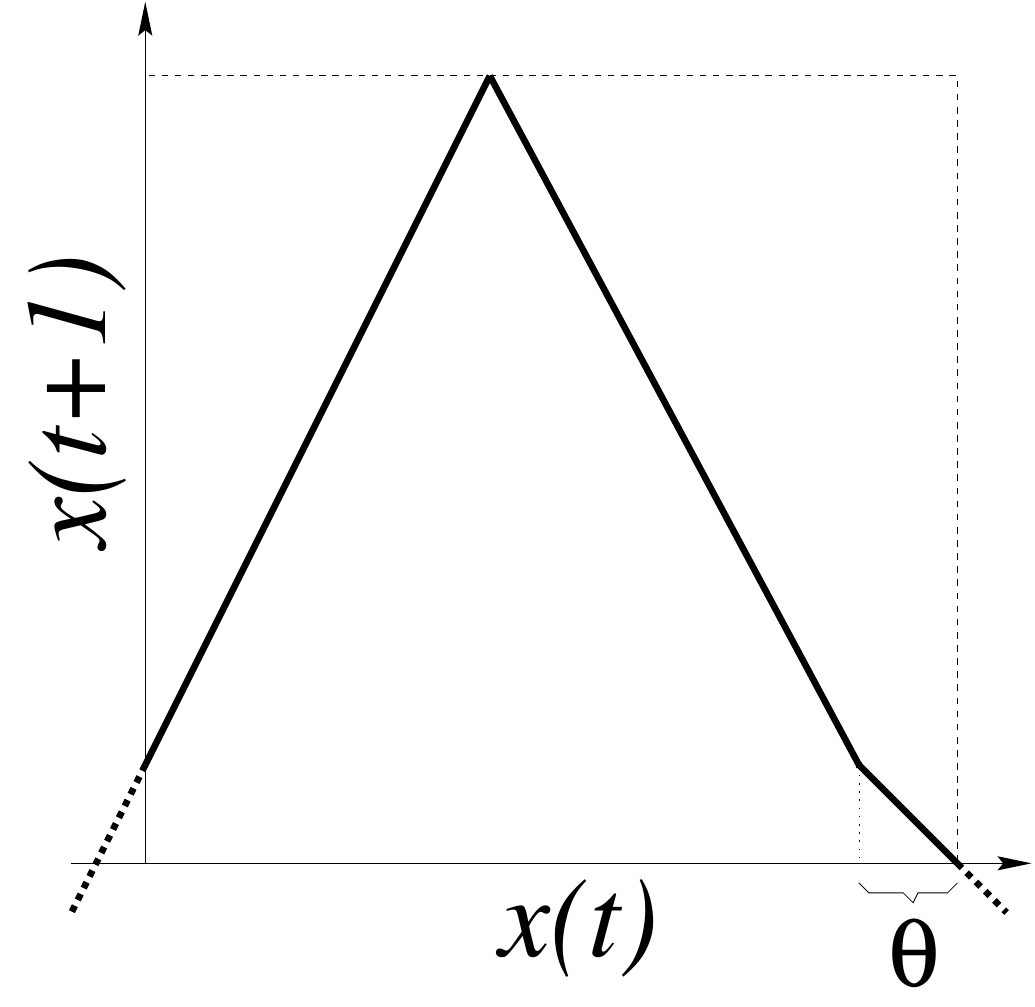}} \hfill
  \raisebox{0.16\linewidth}{%
    \parbox{0.65\linewidth}{\small%
      $x(t+1)=$\\
      \hspace*{0pt}\quad$1-2|x(t)-(1-\theta)/2|+$\\
      \hspace*{0pt}\qquad$\max(x-1+\theta,0)$\\[1ex]
      $\theta\in [0,1/2]$}}
  \caption{%
    \label{fig:ttm}%
    The Tailed Tent Map.}
\end{figure}

Herein, we propose a system based on the most robust map among those
considered in \cite{Callegari:NOLTA-1998} ---~the Tailed Tent Map (TTM)
(Figure~\ref{fig:ttm})%
\cite{Callegari:ISCAS-1997}~--- coupled with one of the analog
memory schemes which best address the signal dependent error issue,
i.e.\@ the \SSI\ technology \cite{Hughes:EL-29-16-1400}. Not surprisingly, this
allows to significantly relax the cycle time limitations and to improve
the performance achievable with respect to previously presented results.

The system is designed with a standard \SI{0.8}{\micro m} \emph{n-well} CMOS
technology and allows an operating frequencies of at least \SI{5}{MHz} ---
which stands out with regard to around \SI{750}{kHz}
in~\cite{Callegari:ISCAS-1997}. This is obtained at a very good level of
esteemed accuracy and with internal currents as low as \SI{\pm 25}{\micro A},
for low-power operation.

\section{Building blocks and interfacing}

The abstract elements illustrated in Figure~\ref{fig:chaosloop} are
herein better investigated to illustrate the adoption of \SSI\ and how
consequent interfacing issues are tackled.

\subsection{%
  \label{sec:TTM}%
  The Nonlinear map}

\begin{figure}[ht]
  \centering
  \resizebox{0.7\linewidth}{!}{\includegraphics{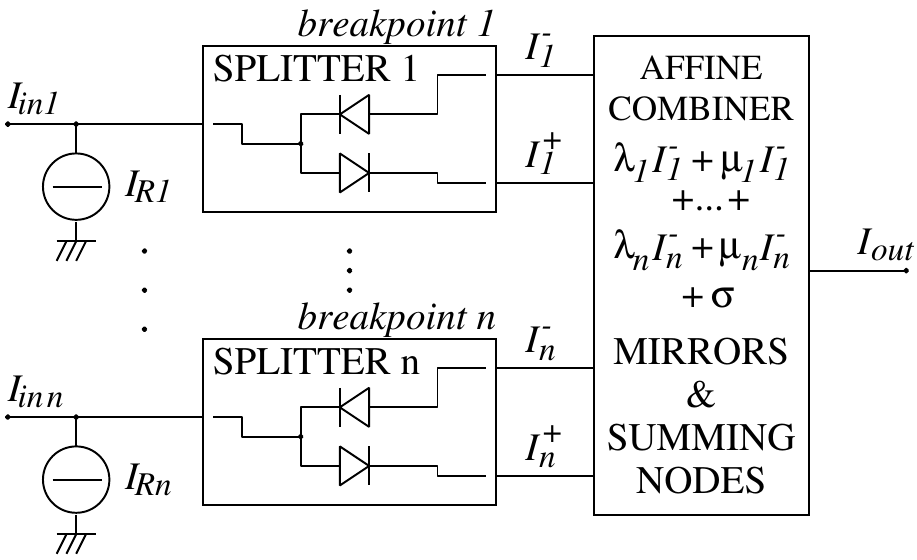}}
  \caption{\label{fig:abstractmap}%
    General PWA current-mode architecture.%
    }
\end{figure}

The most common choice for designing PWA maps is the current mode
approach. Figure~\ref{fig:abstractmap} shows a framework in
which \emph{current splitters} are used to introduce breakpoints by
operating as unidirectional devices. Splitters are arranged to operate
in parallel to reduce propagation times.%

The affine-combiner is implemented by means of mirrors and summing
nodes. A splitter is required for every breakpoint, so that the TTM
requires two of them (and two input currents).

\begin{figure}[ht]
  \centering
  \resizebox{0.6\linewidth}{!}{\includegraphics{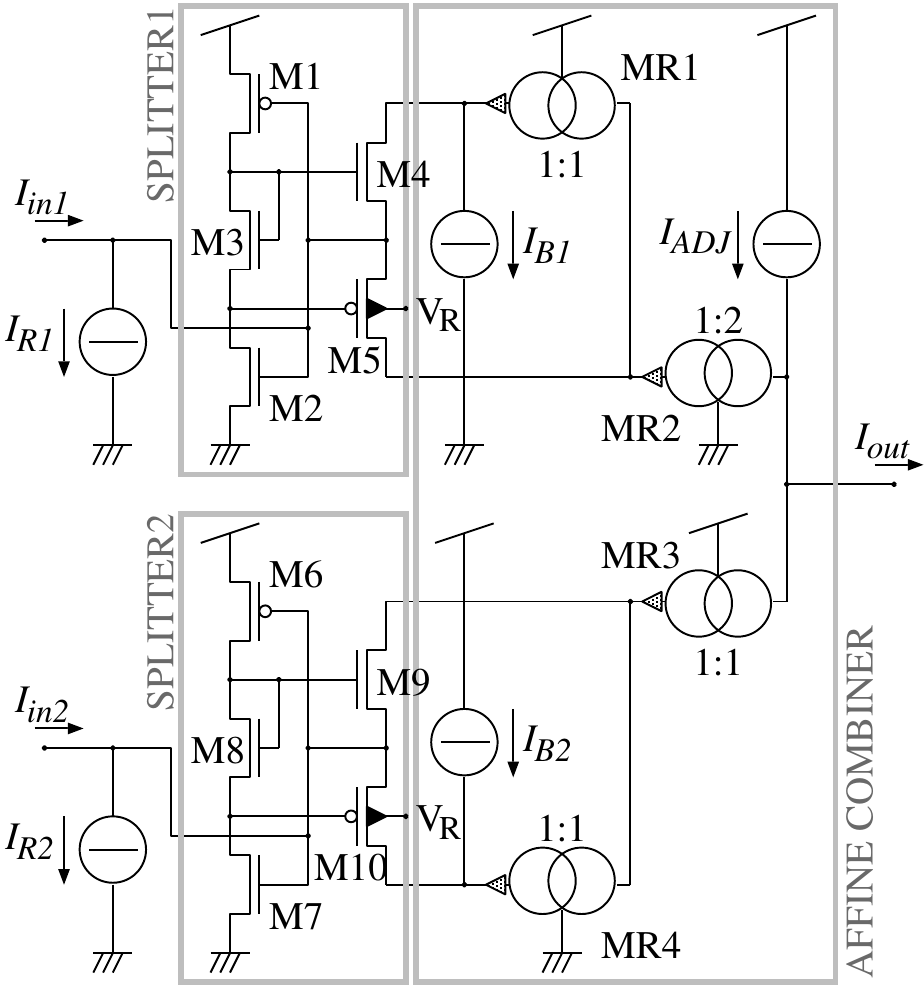}}
  \caption{\label{fig:ttm-circuit}%
    The Tailed Tent Map circuit.}
\end{figure}

The splitters/combiner pair is a potential bottleneck for high
frequency operation since mirror
transistors cannot be taken minimal size, as this would deteriorate
their matching properties and the mirrors accuracy \cite{Pelgrom:JSSC-24-5-1433}. As
a consequence, the
extra-capacitance that has to be driven dictates very careful design.
The proposed TTM circuit (Figure~\ref{fig:ttm-circuit})%
follows the current trend of adopting active splitters
\cite{Delgado:EL-1993-25-2190}. Exploiting an amplifier (M1-M2-M3 or M6-M7-M8),
rather than simply adopting diodes or current mirrors \cite{Langlois:NDES-1995}
reduces the splitter input impedance, speeding up the system response
which is commonly dominated by the product between input impedance and
input capacitance.  With regard to established designs (see
e.g. \cite{Delgado:EL-1993-25-2190}), three modifications are proposed:
\begin{compactenumerate}
\item \textbf{the adoption of transistors} M3 and M8, bringing the input
  impedance further down by assuring that the voltage swing at the input
  capacitance is never larger than
  \begin{equation}
    \label{eq:swing}
    \frac{V_{T_n}+\Veff[N]+|V_{Tp}|+|\Veff[P]|-V_{T_D}-\Veff[D]}{k}
  \end{equation}
  where $V_{T_n}$ is the threshold voltage of M4 (or M9), $\Veff[N]$ is
  its driving voltage, $V_{T_p}$ is the threshold voltage of M5 (or M10),
  $\Veff[P]$ is its driving voltage, $V_{T_D}$ is the threshold voltage
  of transistor M3 (or M8), $\Veff[D]$ is the diode driving voltage and $k$
  is the gain of the inverting amplifier M1-M2-M3 (or M6-M7-M8);
\item \textbf{the adoption of a dummy splitter} driven by no current
  (Figure~\ref{fig:dummy})%
  \begin{figure}[ht]
    \centering
    \resizebox{0.3\linewidth}{!}{\includegraphics{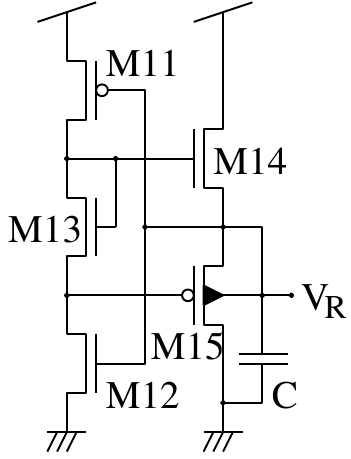}}
    \caption{%
      \label{fig:dummy}%
      The dummy splitter used to generate $V_R$.}
  \end{figure}
  in order to generate a reference voltage $V_R$ to be applied to the
  well of the PMOS transistors M5 and M10.  This avoids body effect,
  reduces $V_{T_p}$ in \eqref{eq:swing}, and loads no extra
  capacitance on the input node as a well-source connection would do.
  If a twin well technology is available, it is convenient to tie also
  the bulks of M4 and M9 to $V_R$;
\item \textbf{the adoption of bias currents}, $I_{B1}$ and $I_{B2}$, pumped
  into mirrors MR1 and MR4 to make sure that in no condition there
  are mirrors copying a null current. In fact, the step response time of
  a current mirror is mostly dependent on the step final current and grows
  very rapidly as this approaches zero.
\end{compactenumerate}

As it will be seen in \S\ref{sec:interfacing}, the search for low
input impedance is not just a matter of speed, but also of interfacing
to \SSI\ stages, so that the extra hardware required for the dummy
splitter is effectively justified. Note also that the mirrors require
cascode structures in order to minimize inaccuracies due to channel
length modulation and finite output impedance. High-swing cascodes are
suggested, as they have a limited hardware cost and they allow the
stacking of the splitters and the mirrors at a power supply of \SI{5}{V}.
For suitable matching, gate areas of mirroring transistors must be no
less than \SIrange{50}{60}{\micro m^2} \cite{Pelgrom:JSSC-24-5-1433}.  Finally,
consider that reference currents $I_{R1}$ and $I_{R2}$ and the offset
cancelling current $I_{\text{ADJ}}$ can be selected among a range of
values, with the result of changing the map invariant set. This degree
of freedom will also be exploited in \S\ref{sec:interfacing}.

\subsection{The \SSI\ analog memory element}  

The memory element required for map iteration must transfer
instantaneously the value at its input to the output when triggered by
a clocking event, and at no time instant is allowed to become
transparent.  To achieve this, it is realized in a master-slave
fashion cascading two sample-and-hold (S/H) circuits driven by
opposite clock phases (balloon in Figure~\ref{fig:chaosloop}).

\begin{figure}[ht]
  \centering \resizebox{!}{2.5cm}{\includegraphics{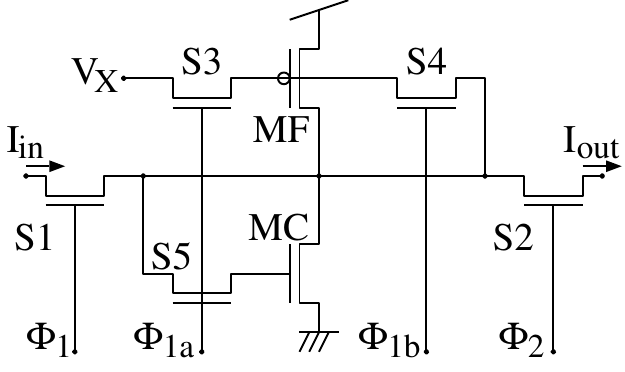}}
  \hspace{0.2cm} \resizebox{!}{2.5cm}{\includegraphics{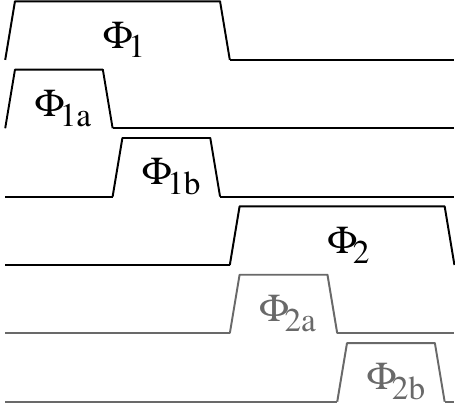}}
  \caption{%
    \label{fig:SSI}%
    Basic \SSI\ circuit and timings. Grayed out clock phases
    $\Phi_{2a}$ and $\Phi_{2b}$ would be used by an identical cascaded
    unit.}
\end{figure}

In its simplest realization, an \SSI\ S/H (or dynamic current mirror)
\cite{Hughes:EL-29-16-1400} is represented in Figure~\ref{fig:SSI}.%
Its working principle is based on a \emph{two step sampling} of the
input current. First of all, MC samples the input current plus a
certain reference current $I_X$ generated by MF, whose gate is
connected to a reference voltage $V_X$ (phase $\Phi_{1a}$). After this
first \emph{coarse} sampling, the current on MC is
$I_{\mathit{in}}+I_X+\epsilon_C$ where $\epsilon_C$ is an error mainly
due to clock feed-through on switch S5. In the second step,
MF samples the difference between the input current and the current
stored on MC, i.e.\@ $I_X+\epsilon_C$ (phase $\Phi_{1b}$).  Phase
$\Phi_2$ is the \emph{hold} phase. Also at the end of the \emph{fine}
sampling $\Phi_{1b}$ an error is committed, so that the final current
stored on MF is $I_X+\epsilon_C-\epsilon_F$. Hence, the output current
of the \SSI\ S/H is $I_{\mathit{in}}+\epsilon_F$.

With this, even though an \SSI\ S/H stage is not immune from sampling
errors, the error $\epsilon_F$ is introduced while sampling a current
always very close to $I_X$ so to be an (almost) signal independent
error.

Moreover, a whole memory element is made of two S/H stages.  In the
case of \SSI\ stages, two identical units can be directly cascaded,
thanks to the presence of the reference current $I_X$ which makes an
\SSI\ S/H capable of dealing with both positive and negative input
currents. This arrangement has the property of cancelling signal
independent errors \cite{Machado:LPHF}, so that the operation of a
\SSI\ 2 stage memory element is (theoretically) error free.

Note that, for cascading, the down-hill unit requires clocking signals
complementary to those used in the up-hill one: in
Figure~\ref{fig:SSI}, signals $\Phi_2$, $\Phi_{2a}$ and $\Phi_{2b}$
are meant to drive the second S/H. Clock phases are rigorously
non-overlapping.

\subsection{%
  \label{sec:interfacing}%
  Interfacing}

This section deals with the problem of interfacing a TTM circuit
(\S\ref{sec:TTM}) to an \SSI\ memory element.

First of all, electrical interfacing will be considered. \SSI\ circuits
do not pose particular requirements to units driving them, but
\emph{do} pose constraints on the circuits being driven by them, since
they are fundamentally designed to be cascaded to other \SSI\ units.
This is evident if one looks at the absence of cascode stages replacing
over MC and MF, which may therefore be affected by channel length
modulation. Errors are prevented by the fact that when an \SSI\ unit is
cascaded to another identical one, the fine sampling phase of the
second unit guarantees the voltage at the drains of MC and MF to be
approximately $V_X$, i.e.\@ the same voltage as at the end of the
first unit sampling.

To prevent channel modulation error when driving a non-\SSI stage
without having to provide cascode structures for MC and MF, we need to
adjust the input voltage of the circuit being driven by the down-hill
\SSI\ stage to be \emph{constantly} as similar as possible to the
$V_X$ used as a reference in the \SSI\ circuits. To do so, a very low
input impedance is needed.

Connecting to conventional mirrors is thus generally not feasible due
to their non-negligible input impedance. What we propose is instead to
connect the \SSI\ memory element directly to the TTM circuit. In fact,
this block has been designed for an extremely low input impedance
(\S\ref{sec:TTM}) and for an input voltage almost fixed at $V_R$. The
dummy splitter already introduced in \S\ref{sec:TTM} is here of great
help as it can be exploited to generate $V_X\equiv V_R$.

\begin{figure}[ht]
  \centering
  \resizebox{\linewidth}{!}{\includegraphics{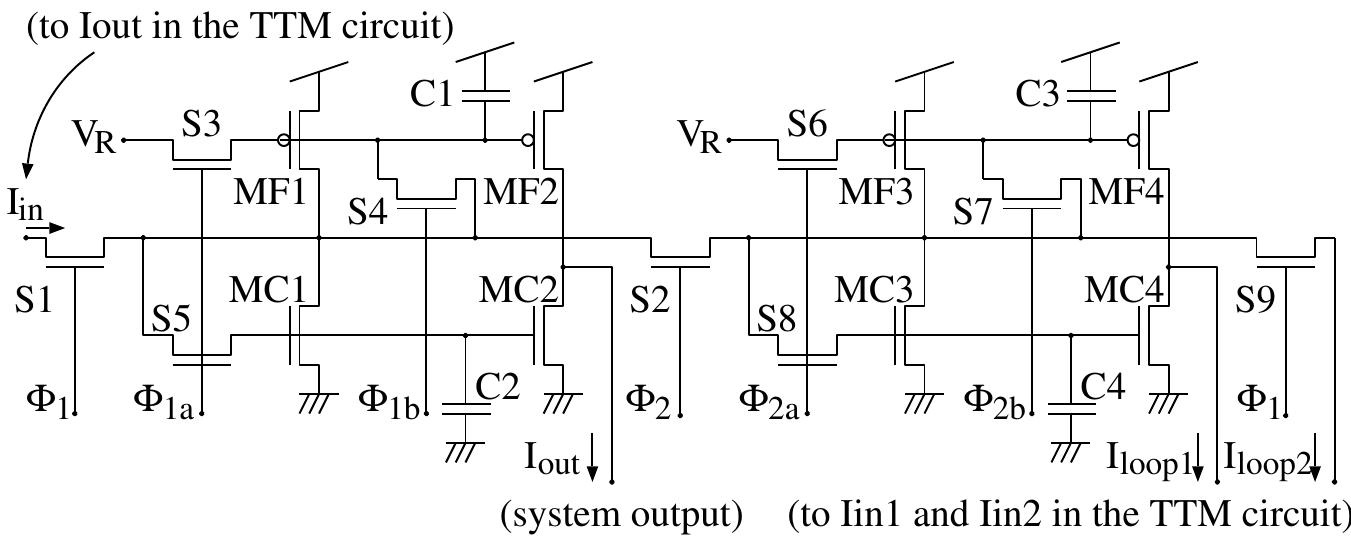}}
  \caption{%
    \label{fig:amem}%
    A \SSI\ memory element that can be directly interfaced with the TTM
    circuit shown in Figure~\ref{fig:ttm-circuit}.}
\end{figure}

The second point is functional interfacing. The TTM circuit requires
two input currents.  Furthermore, an additional current is necessary to
read the output of the overall chaotic system.
To cope with this, we propose to build current mirrors \emph{inside}
the \SSI\ memory element. The final, complete circuit is
shown in Figure~\ref{fig:amem}.%
Such a circuit can be connected to the TTM circuit exploiting input
$I_{\text{in}}$ and outputs $I_{\text{loop1}}$ and $I_{\text{loop2}}$.
$I_{\text{out}}$ is for reading the overall chaotic system output: it offers a
time shifted (half clock cycle), complemented version of the system state
variable. In the whole circuit this is the only signal subject to a
non-compensated sampling error (which happens to be a constant offset not
particularly troublesome for most applications).  Note that mirroring inside
the \SSI\ subsystem introduces matching errors (which would anyway be
introduced elsewhere in the signal path). To keep them low, all the MC\emph{x}
and the MF\emph{x} transistors must be sized to have an area greater than
\SIrange{50}{60}{\micro m^2}. The consequence is a noticeable sampling error
due to charge feed-through on the drain-to-gate capacitances of MC\emph{x} and
the MF\emph{x}. Nonetheless proper device dimensioning can keep such error
tolerable as confirmed by the the results in \S\ref{sec:simula}.

As a further consideration, note that sampling error may also be
reduced by exploiting the ability of \SSI\ S/H to cope with either
positive or negative inputs and adopting a signal range centered on
zero. In fact, this allows to use the smallest possible reference
currents, to which errors are tied.

To do so, the values of $I_{R1}$, $I_{R2}$ and $I_{\text{ADJ}}$ can
be chosen so that the system invariant set is centered around
zero. The circuit has been simulated with $I_{R1}= \SI{-1.25}{\micro A}$,
$I_{R2}=\SI{22.5}{\micro A}$, $I_{B1}=I_{B2}=\SI{2.5}{\micro A}$ and
$I_{\text{ADJ}}=\SI{32.5}{\micro A}$, so that the ``state variable''
currents are in the \SI{\pm25}{\micro A} range.

\begin{figure}[ht]
  \centering
  \resizebox{\linewidth}{2cm}{\includegraphics{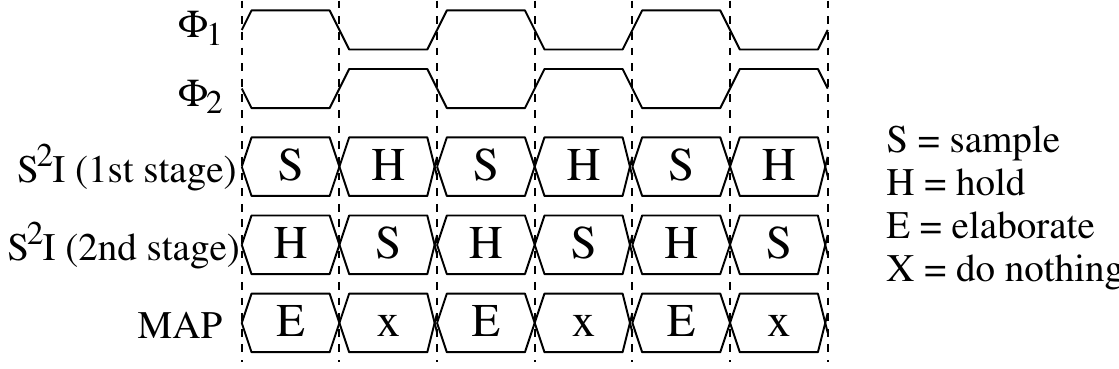}}
  \caption{%
    \label{fig:timing}%
    Basic timings of the overall TTM+\SSI\ system.}
\end{figure}

Finally, we shall briefly consider timings. Being the \SSI\ dynamic
mirror a sample-and-hold rather than a track-and-hold unit, it cannot
deliver output while sampling. Hence, only half of clock cycle is left
to the map circuit to elaborate its output (see
Figure~\ref{fig:timing}).  While this poses constraints on the
response time of the map circuit, it might (at least theoretically)
allow interleaving of two analog memory elements on the same TTM
circuit.

\section{%
  \label{sec:simula}% 
System simulation and conclusions}

\begin{figure}[ht]
  \centering 
  \resizebox{0.8\linewidth}{3.5cm}{\includegraphics{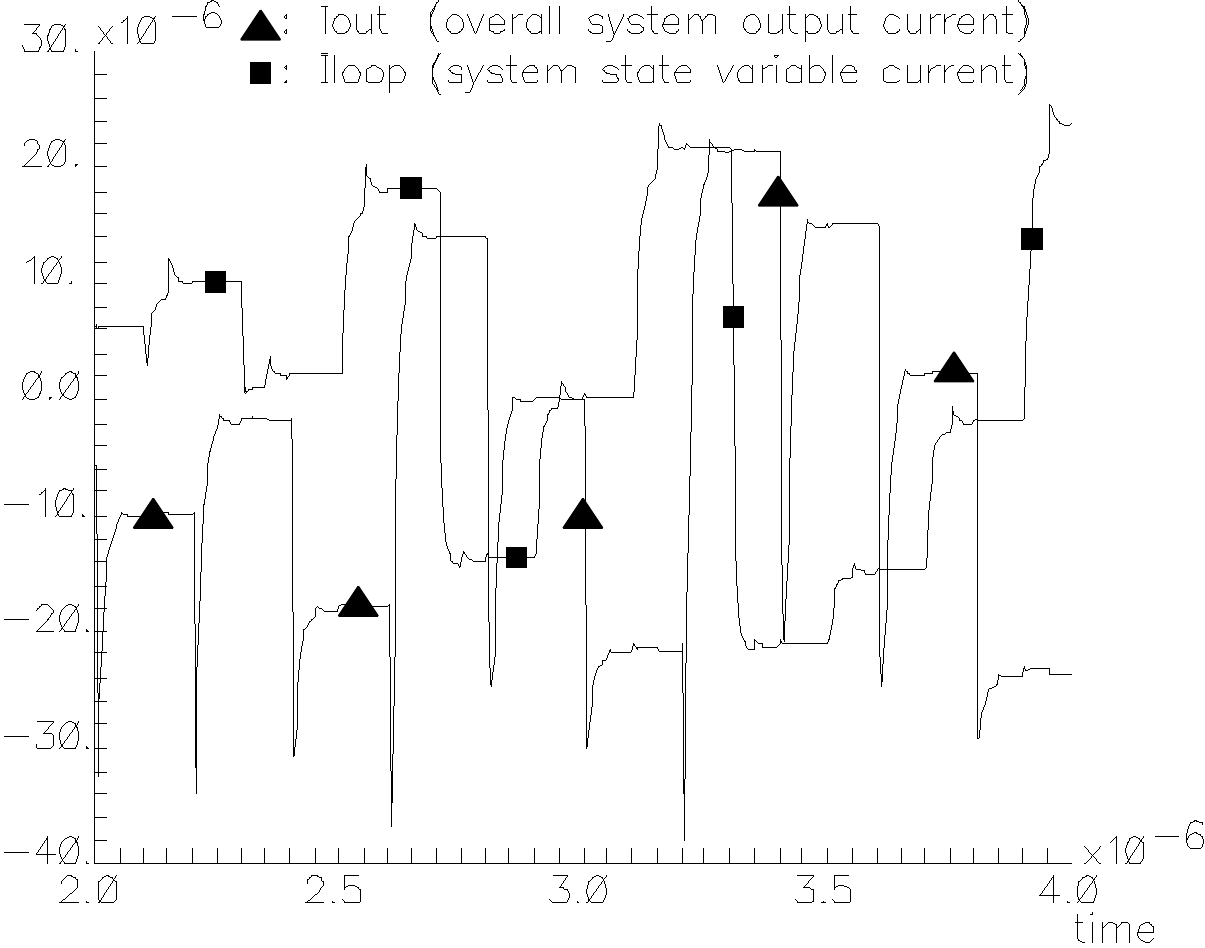}}
  \caption{%
    \label{fig:sim}%
     A chaotic sequence as obtained by the overall system. Both the output
     current and the ``state variable'' current are shown.}
\end{figure}

\begin{figure}[ht]
  \centering
  \resizebox{0.8\linewidth}{4cm}{\includegraphics{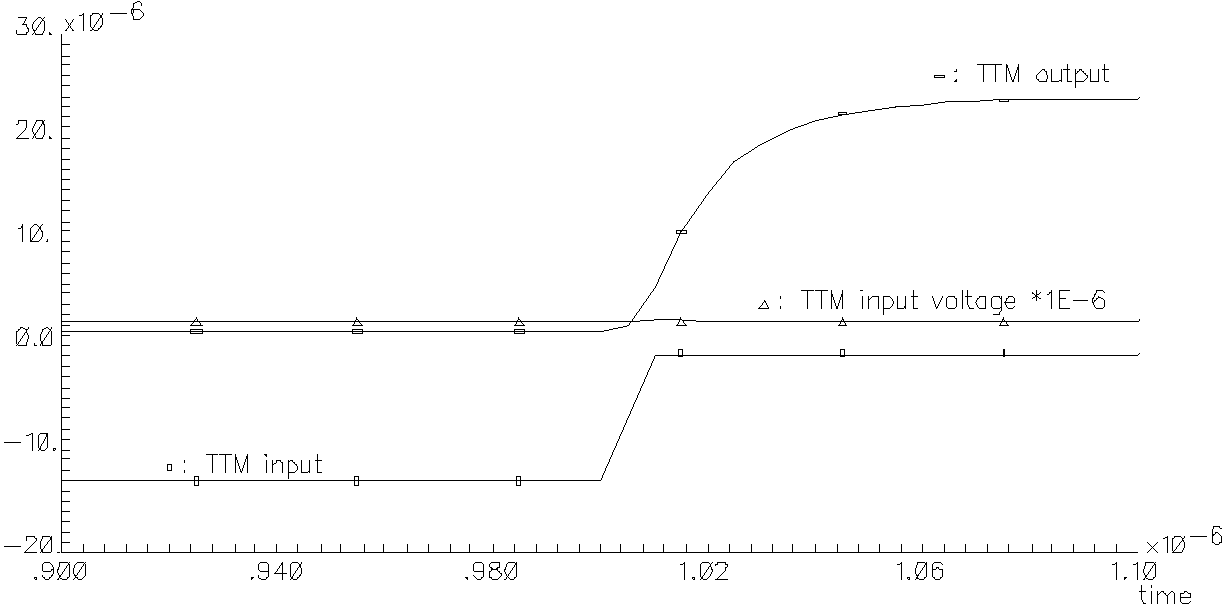}} 
  \caption{%
    \label{fig:sim2}%
    Worst case switching of the TTM circuit. Traces show the input
    current, the output current and the voltage at one of the TTM
    inputs, confirming the low input impedance of the circuit.}
\end{figure}

System simulations were run using SPECTRE, referring to
charge-conservative BSIM3 device models.  Figure~\ref{fig:sim}%
shows a collection of output samples as produced by the overall system. The
operating frequency of \SI{5}{MHz} is limited by the worst-case step response
time of the TTM circuit (about \SI{90}{nS} --- Figure~\ref{fig:sim2}) which
corresponds to a step in the input current taking a splitter right over or
below threshold.

\begin{figure}[ht]
  \centering
  \resizebox{0.7\linewidth}{3.5cm}{\includegraphics{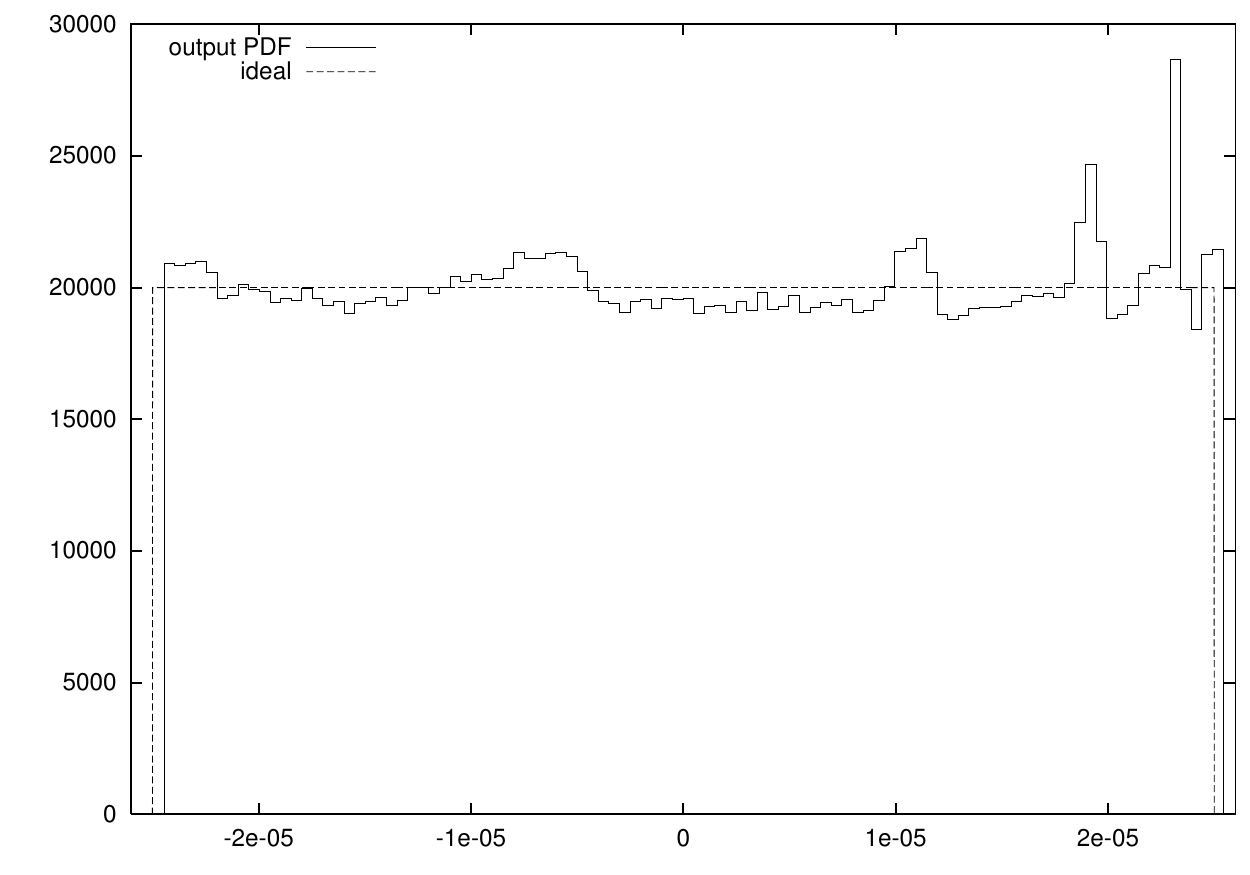}} 
  \caption{%
    \label{fig:PDF}%
    PDF of the system simulated using \emph{C} code and an
    approximation of the open-loop behavior. Offsets are due to the
    fact that the output is subject to signal independent errors.}
\end{figure}

To obtain a reliable estimation of the asymptotic statistical distribution of
the system states in a reasonable amount of time, we have simulated the
behavior of an open-loop system based on the TTM circuit together with the
\SSI\ subsystem, driven by a stepping signal spanning the whole range of
operation of the TTM circuit (.TRAN analysis). Successively, the collected
information has been used to build and interpolating behavioural model in
\emph{C} code. Finally, this efficient model has been iterated to produce the
large number of samples required for PDF estimation.  The result is reported in
Figure~\ref{fig:PDF} to show a good agreement with the theoretical
expectations.

\small
\bibliographystyle{ieeetr}
\bibliography{chaos_theory,pram,chaos_circuits,tent_map,analog} 
%\bibliography{chaotic,pram,tent,analog,chaos_theory,chaos_circuits,tent_map}
\end{document}